\documentclass[showpacs,preprint,aps]{revtex4}

\usepackage{epsfig}
\usepackage{dcolumn}
\usepackage{bm}
\usepackage{amsmath}

\def\vr{\bm{r}}
\def\D{\frac{\hbar^2}{2m}}
\def\Dd{\frac{\hbar^2}{4m}}
\def\DD{\frac{\hbar^2}{8m}}
\def\g{\gamma}
\def\R{\rho}

\begin{document}

\title{Density functional study of two--dimensional $^4$He clusters}
\author{J. Mur-Petit}
\affiliation{Dept.\ d'Estructura i Constituents de la Mat\`eria,
Universitat de Barcelona, Avda. Diagonal, 647, E-08028 Barcelona, Spain}
\author{A. Sarsa}
\affiliation{Depto. de F\'{\i}sica, Campus de Rabanales, 
Universidad de C\'ordoba, E-14071 C\'ordoba, Spain }
\author{J. Navarro}
\affiliation{IFIC, CSIC-Universitat de Val\`encia, Apdo. 20285,
E-46071 Val\`encia, Spain}
\author{A. Polls}
\affiliation{Dept.\ d'Estructura i Constituents de la Mat\`eria,
Universitat de Barcelona, Diagonal, 647, E-08028 Barcelona, Spain}

\date{\today}

\begin{abstract}
Binding energies and density profiles of two-dimensional systems of liquid
$^4$He with different geometries are studied by means of a zero-range 
density functional adjusted to 
reproduce the line tension obtained in a previous
diffusion Monte Carlo calculation ($\lambda_{DMC}=0.121$ K/\AA).
It is shown that this density functional provides accurate results for the 
binding energy of large clusters with a reasonable computational effort.
\end{abstract}

\maketitle

\section{Introduction}
Quantum liquids in restricted geometries have attracted a lot of attention
in recent years.~\cite{kro02} One interesting feature of these systems is that
their internal structure is more  accesible than in bulk liquids due to
the restricted motion of the atoms in the confining potential.  
Among these systems the study of two-dimensional systems has received
particular attention. An example of such a system is liquid helium adsorbed to
a more-or-less attractive flat surface. 
This kind of system was observed for the first time by M. Bretz {\it et
  al.}~\cite{bre73} in 1973, when they reported the observation of
adsorbed $^4$He onto the basal plane of graphite. In the last few years,
adsorption properties of helium on many different substrates (carbon, alkali
and alkaline-earth flat surfaces, carbon nanotubes, aerogels) have become a
fruitful topic of research.

Theoretical microscopic studies with realistic atom-atom interactions, such as
that of Clements {\it et   al.}~\cite{cle93}, have shown that films with low
surface coverages, where all atoms cover the surface with a thickness
corresponding to a single atom, can be reasonably approximated by a
two-dimensional model. In connection with these systems, an interesting
question naturally arises as how physics depends on the dimensionality of the
space.

The homogeneous 2D liquid has been studied using different theoretical
methods, such as molecular dynamics~\cite{cam71} and quantum Monte
Carlo simulations either Green's Function~\cite{whi88} or
diffusion~\cite{gio96} techniques. 

Recently, two-dimensional clusters of liquid 
$^4$He  have been studied using 
using a shadow variational wave function~\cite{kri99}, and also by 
the diffusion Monte Carlo (DMC) method~\cite{old}. In these two references,
the binding energies of the 2D-clusters were fitted by means of a mass formula 
and a line tension of $\lambda = 0.121 $ was reported in Ref. \cite{old}.
However, due to computational limitations, the number of atoms 
in the clusters was limited to $N \sim 100$. 
Besides, the density profiles of the clusters, 
specially for coordinates close to the origin, are usually obtained in 
the Monte Carlo method with poor statistics. 
Therefore it seems appropiate to build a density functional suitable 
to be used in 2D, using the same procedure which has already been 
succesfully used to study 3D $^4$He clusters \cite{str87,barran93}.

Density  functionals are based on the well-known Hohenberg-Kohn theorem~\cite{hk}
that asserts that the ground-state energy per particle of a many-body system
can be written as a functional of the density.
Once the functional is available, its minimization  
brings to an Euler-Lagrange equation for the density profile $\rho(\vr)$,
which allows to calculate the properties of clusters.  
Note that the results obtained with the density functional 
will be more reliable for larger clusters, as it has been constructed to 
reproduce properties of the homogeneous and semi-infinite media.

The density functional we use is the simplest version of the
zero-range functional intensively   used in 3D~\cite{strin2}. 
Its parameters have been  
adjusted so as to reproduce some properties of the ground state of the 
homogeneous system as obtained in DMC calculations~\cite{gio96}, as
well as the line tension extracted from the mass formula of Ref.~\cite{old}.  
This procedure is discussed in Sect.~\ref{ss:sl}, together with the results
for the slabs. Sect.~\ref{ss:dr} is devoted to the study of finite droplets,
with special emphasis for those with a large number of atoms.
Finally, in Section~\ref{s:conc} the main conclusions are summarized.

\section{Semi-infinite system and slabs}
\label{ss:sl}
Density functionals  to investigate surface properties of superfluid $^4$He
were developed during the 1970's \cite{ebner}. At zero temperature and in 
the absence of currents, the order parameter of a bosonic system is nothing 
but the square root of the one-body density $\R(\vr)$, so it is natural to 
think of the energy of the
system as a functional of the helium density. Stringari~\cite{strin2} proposed
a zero-range density functional  for non-homogeneous three-dimensional $^4$He 
systems of the form:
\begin{eqnarray}
  E[\R] &=& 
   \int\,d\vr\left\{\D \frac{|\nabla\R|^2}{4\R}
   +b\R^2+c\R^{2+\g}+d|\nabla\R|^2\right\},
   \label{eq:zr-func}
\end{eqnarray}
which  was  used to study $^4$He surface properties 
\cite{strin3} and clusters \cite{str87}.

For the two-dimensional systems, we will use the same functional form. 
However, the parameters $b,c,\g$ characterizing the functional for 
homogeneous systems have to be recalculated by 
requiring that the functional reproduces some known properties of
the two-dimensional $^4$He system, such as the ground-state energy
per particle ($e_0=-0.89706$ K), saturation density ($\R_0=0.04344$
\AA$^{-2}$), and the compressibility (or the velocity of sound, $s=92.8$
m/s), which were derived in the framework of the Difusion Monte 
Carlo method~\cite{gio96}. 
In addition, the parameter $d$ is obtained by demanding that the 
line tension of the semiinfinite system equals that obtained from 
DMC calculations for 2D clusters, $\lambda=0.121$ K/\AA~\cite{old}.
The energy per particle for the homogeneous system provided by this functional 
 reproduces very well the equation of state
obtained from DMC calcualtions in a wide range of densities \cite{gio96}.  
Therefore, as a first step we need to study the semi-infinite system. 
In principle, one should solve an Euler-Lagrange equation for the
density profile $\R(\vr)$ which results from minimizing the energy 
functional of Eq.~(\ref{eq:zr-func}) after introducing the 
Lagrange multiplier ($\mu$), which is identified with the 
chemical potential, 
\begin{equation}
  \DD\left[-\frac{2\nabla^2\R}{\R}+\frac{|\nabla\R|^2}{\R^2}\right]
    +2b\R+(2+\g)c\R^{1+\g}-2d\nabla^2\R = \mu \;.
  \label{eq:zr-EL}
\end{equation}
One should also impose the boundary condition, 
$\rho(x \rightarrow -\infty)= \rho_0$.
However, for the particular case of a zero range functional, 
the line tension of the semi-infinite system 
 can be evaluated in a closed form, without solving for the density profile:
\begin{equation}
  \lambda = 2\int_0^{\R_0}\;\mathrm{d}\R\left[
    \left( \DD+\R d\right)
    \left( b\R+c\R^{1+\g}-\mu \right) \right]^{1/2} \;.
  \label{eq:zr-lambda}
\end{equation}
Then, imposing $\lambda=0.121$ K/\AA~\cite{old}
one  gets an implicit equation $\lambda=\lambda(d)$ which can be solved
numerically, thus fixing the last parameter of the functional.
The parameters for the two-dimensional zero range functional are
listed in Table~\ref{t:func}. 

Once the functional is defined, one can
study any two-dimensional $^4$He system.
 The first systems we have considered are
two-dimensional slabs with varying central density $\R_c= \R(0)$. 
A systematic study of three-dimensional $^4$He slabs with different 
density functionals has been presented in Ref. \cite{szy}
The Euler equation for the slabs is the same as for the 
semi-infinite system (Eq.~(\ref{eq:zr-EL})).
The changes in the solution of the equation originate from the 
different geometry and boundary conditions which define the slab. 
Translational symmetry implies that the density depends only on the 
coordinate perpendicular
to the slab surface, which we call $x$. Then, $\nabla\R=\R'(x)$ and
$\nabla^2\R=\R''(x)$, where the prime denotes derivative with respect to $x$.
In this way, the Euler equation (Eq. (\ref{eq:zr-EL})) can be expressed in 
a more convenient form, in which now $\R$ depends only on  $x$,
\begin{equation} 
\frac {\hbar^2}{8m} \left ( \frac {(\R')^2}{\R^2} - \frac {2  \R''}{\R}
\right ) - 2 d  \R'' + 2 b \R + (\gamma+2) c \R^{1+\gamma} = \mu
\label{eq:euler2}
\end{equation}
In  the next step, one eliminates the second derivative of 
Eq. (\ref{eq:euler2}) by multiplying  both sides of the equation  by 
$\R'$ and integrating with respect to $x$, from the origin to a given 
value of $x$, 
\begin{equation}
\left [ -\frac {\hbar^2}{8m} \frac {(\R')^2}{\R} - d (\R')^2 + b \R^2 +
c \R^{\gamma+2} \right ]_0^x = \mu \left [ \R(x) - \R(0) \right ].
\label{eq:euler3}
\end{equation}
Imposing $\R(\infty)= \R'(\infty)=0$, and 
considering that the slab  
is symmetric respect to $x=0$, and therefore $\R'(0)=0$, 
one obtains  the chemical potential as a function of 
the central density of the slab,
\begin{equation}
\mu = b \R_c + c \R_c^{\gamma +1} .
\label{eq:chem}
\end{equation}
We remark that this chemical potential is constant along the profile. 

Going back to Eq. (\ref{eq:euler3}),
and using  the fact that  $\R'(0)=0$, one obtains

\begin{equation}
 \R' = \sqrt{\frac{\R^2 (b\R + c\R^{\gamma+1} -\mu )}{ \hbar^2 /8 m + d \R} }.
\label{eq:euler4}
\end{equation}
This expression for $\R'$ is then used to calculate the number of 
atoms per unit length along the $y$ axis (parallel to the surface),
i.e. the coverage, in a closed expression in terms of $\R$,
\begin{equation}
\frac {N}{L} = 2 \int_0^{\infty} \R(x) \mathrm{d}x = 2 \int_0^{\R(0)} \R 
\frac {\mathrm{d} \R}{\mid \R' \mid}=
2\int_0^{\R (0)} \mathrm{d}\R \sqrt{\frac{\hbar^2/8m + d\R}{b\R + c\R^{1+\gamma} -\mu}}.
\end{equation}
This integral has a singularity when $\R \rightarrow \R_c $, 
which can be avoided performing and integration by parts. 
The final expression, free of numerical problems reads 
\begin{multline}
\frac {N}{L} = -\frac {4}{b} \sqrt {- \frac {\hbar^2}{8m} \mu } \nonumber\\
- 4 \int_0^{\R(0)} \mathrm{d}\R
\frac {(b\R +c \R^{1+\gamma} -\mu)^{1/2} 
  \left [ \frac {d}{2} (b +c(1+\gamma) \R^{\gamma} )
    - (\frac {\hbar^2}{8m} + d \R) c \gamma (1+\gamma) \R^{\gamma-1} \right ] }
{(b +c (1 +\gamma) \R^{\gamma})^2 (\frac {\hbar^2}{8 m} + d \R )^{1/2} }
\label{eq:cover}
\end{multline}

Also useful is the energy per unit length,
\begin{equation}
  \frac{E}{L} = 2\int_0^{\infty} \mathrm{d}x \left[\frac{\hbar^2}{8m}\frac{(\R')^2}{\R} 
    +b \R^2 + c \R^{2+\gamma} + d (\R')^2 \right ].
\end{equation}
Using the Euler equation (Eq. \ref{eq:euler4}) and the previous definition of 
the coverage, one can finally express the energy per particle $e = E/N$ in 
terms of the inverse of the coverage, $\tilde x=L/N$,
\begin{equation}
e(\tilde x) = \mu(\tilde x) + 4 \tilde x \int_0^{\R(0)} \mathrm{d} \R 
\sqrt {\left(\frac {\hbar^2}{8m} + d \R\right) (b \R + c \R^{1+\gamma} - \mu(\tilde x))}  
\label{eq:ener1}
\end{equation}
Therefore, given $\R(0) \in [0, \R_0]$, and using the previous expressions, 
one can calculate the chemical potential, the coverage and the energy 
per particle. 

\begin{figure}[t]
\includegraphics*[width=12cm]{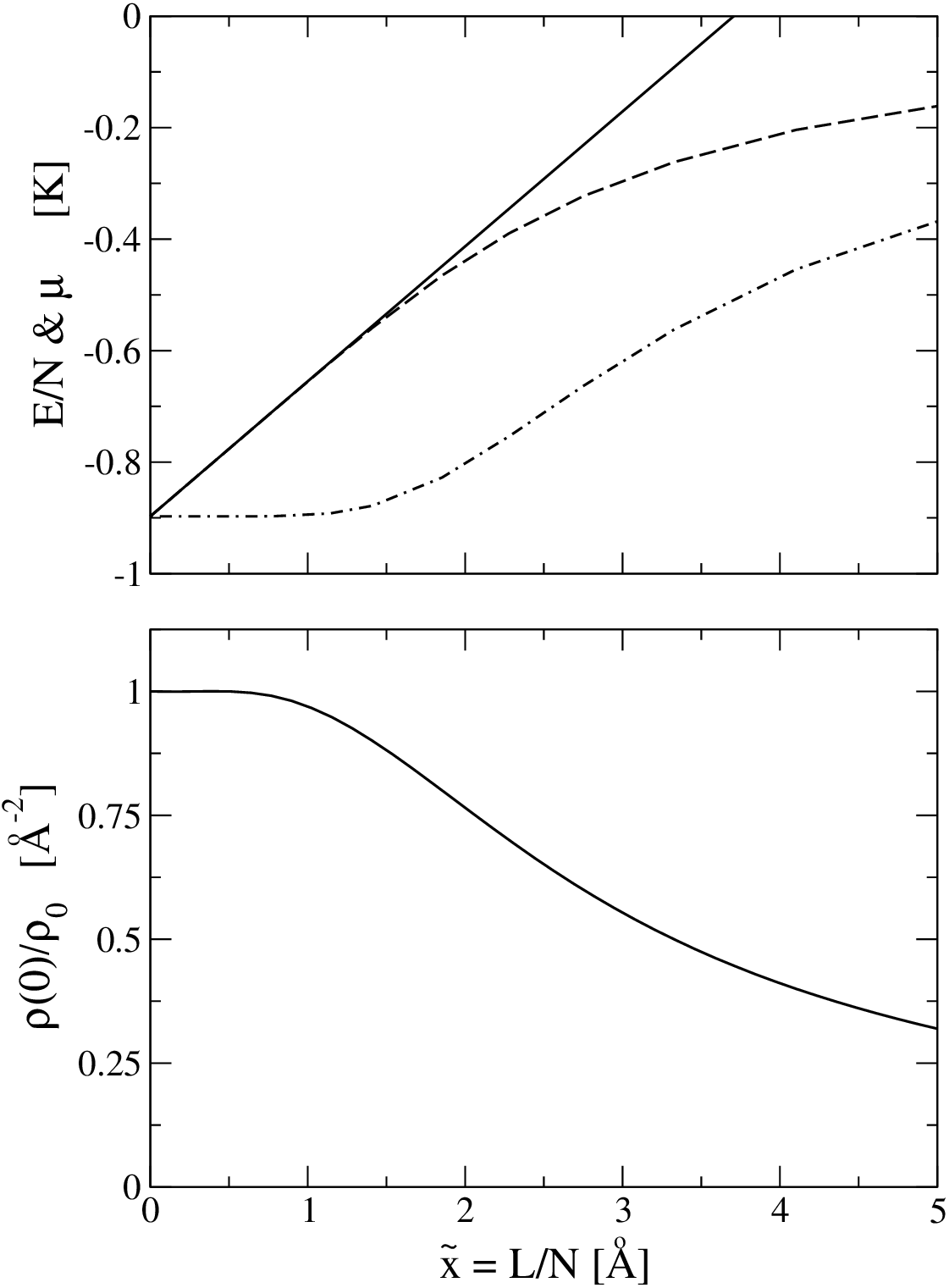}
\caption{(top) Energy per particle (dashed line) and chemical potential 
(dot-dashed line) for $^4$He slabs as a function of the  inverse of the 
coverage. The solid straight line corresponds to the asymptotic behavior 
of the energy per particle and its slope is determined by the linear 
tension of the semiinfinite system.
(bottom) Ratio between the central density  and the bulk 
equilibrium density  for $^4$He slabs as a function of the 
inverse of the coverage.}
\label{fig:ener-slab}
\end{figure}

The energy per particle and the chemical potential as a function 
of the inverse of the coverage $\tilde x$ are reported in 
Fig.~(\ref{fig:ener-slab}). In the limit $\tilde x \rightarrow 0$ 
one recovers the binding energy at $\R_0$ of the uniform system,
which in turn coincides with the chemical potential. The energy per particle
has a very clean linear behavior at the origin as illustrated by the 
solid straight line which provides a very good description of the energy 
per particle up to $\tilde x \sim 1.5$ \AA. The slope of this line can 
be analytically derived and turns out  to be twice the linear tension. 
As a consequence of the linear tension, the binding energy per particle 
of the slab decreases with the inverse of the coverage. 
Actually, the derivation of the linear behavior of the binding energy per 
particle can be obtained starting from  Eq. (\ref{eq:ener1}), which defines 
the energy per particle as a function of $\tilde x$, by performing an expansion
around $\tilde x=0$,  
\begin{equation}
  e(\tilde x) = \mu_{\infty} + 2 \sigma \tilde x + ...
\label{eq:expa}
\end{equation}

For values larger than $\tilde x\sim 1.5$ \AA\ the binding energy 
per particle starts to bend towards the $\tilde x$ axis and becomes a 
convex function, which will slowly approach zero.
The chemical potential is very flat at the origin being  determined by the 
central density of the slab. The ratio of the  central density to 
the equilibrium density  as a function of the inverse
of the coverage is displayed in the lower panel of 
Fig.~(\ref{fig:ener-slab}). In agreement with 
the chemical potential, the central density is very flat at small values 
of $\tilde x$. The central density of a slab  can never go above the 
equilibrium density and it is always a decreasing function of $\tilde x$. 
The flatness of the central density and the chemical potential for small 
values of $\tilde x$, indicates 
that the slab approaches very slowly the limit of the infinite system.

\begin{figure}[t]
\includegraphics*[width=12cm]{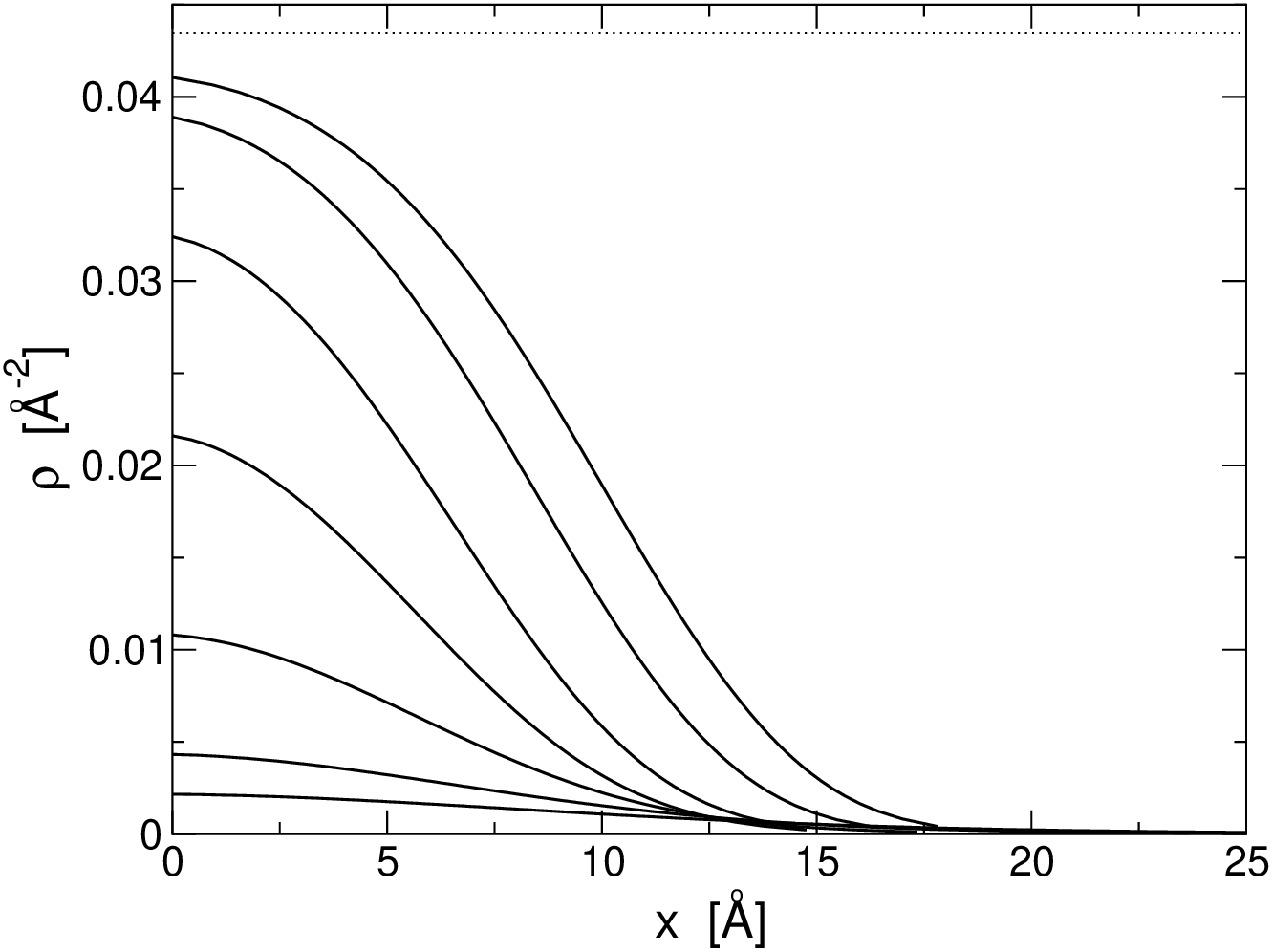}
\caption{Density profiles for $^4$He slabs with central densities 
$\R_c/\R_0$=0.05, 0.1, 0.25, 0.5, 0.75, 0.9, 0.95, where $\R_0$ is the 
equilibrium density of the bulk. The horizontal dotted line corresponds to the
bulk density $\rho_0$.} 
\label{fig:prof-slabs}
\end{figure}

\begin{figure}[t]
\includegraphics*[width=12cm]{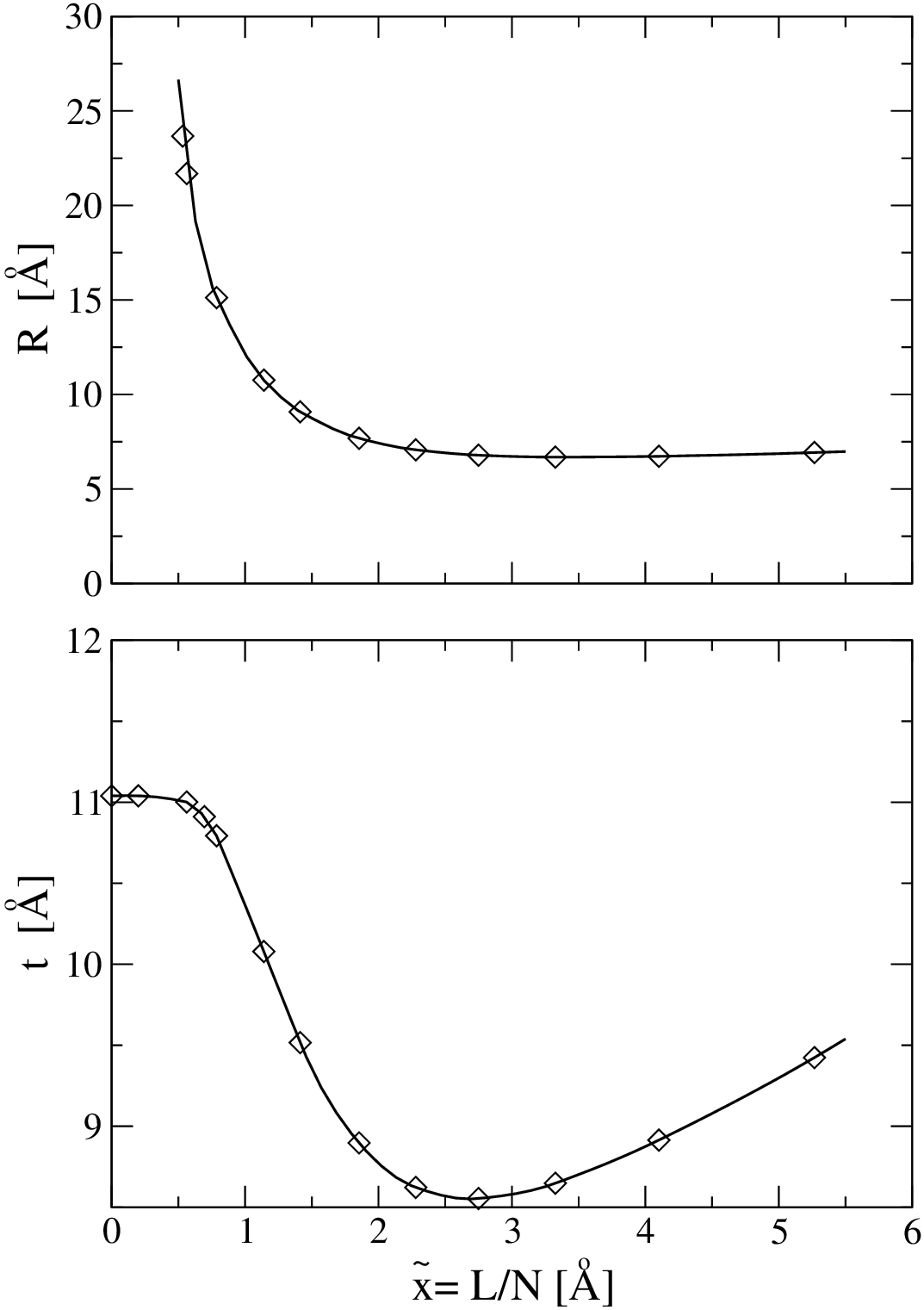}
\caption{Radius (top panel) and thickness (lower panel) of $^4$He slabs as 
a function of $\tilde x$. The symbols are the calculated data while the lines
are cubic splines to guide the eye.} 
\label{fig:radius.slab}
\end{figure}

The density profiles of the slabs
can be obtained from the following relation
\begin{equation}
  \int_{\R(x)}^{\R_c} \mathrm{d}\R \left[
    \frac{\DD\frac{1}{\R}+d} {b\R^2+c\R^{2+\g}-\mu\R}\right]^{1/2} = x\;,
  \label{eq:zr-profile}
\end{equation}
valid for $x \ge 0$. Notice also the presence of a divergency when 
$\rho \rightarrow \rho_c$ which can be again avoided performing 
an integration by parts.
The profiles calculated for various central densities are plotted in
Fig.~(\ref{fig:prof-slabs}). The size of the slab increases with the
central density. A measure of this size is given by the radius $R$, defined 
as the distance from  the origin to the point where the density has 
fallen to half its central value. The radius of the slabs as a function 
of $\tilde x$ is shown in the top panel of Fig.~(\ref{fig:radius.slab}). 
As expected, the radius diverges in the limit $\rho_c \rightarrow \rho_0$ 
and is a decreasing function of $\tilde x$.
It also presents a very shallow minimum around $\tilde x \sim 3.3$ \AA. 
The profiles are also characterized by the thickness $t$, 
defined as the distance between the points where the density 
has decreased from $90 \%$ to $10 \%$ of its central value. 
The thickness as a function of $\tilde x$ is reported in 
the lower part of Fig.~(\ref{fig:radius.slab}). 
Up to $\tilde x\sim 1$, the thickness
is a flat function, indicating that the surface of the slab is 
very much the same, actually is the radius of the slab that grows very 
fast and diverges when $x \rightarrow 0$.  
The thickness is very large for the smaller slabs: as each particle
interacts with very few others, the system extends to very large distances.
When the central density is increased ($\tilde x$ decreases), the
thickness decreases until it has a minimum at $\tilde x\approx 2.75$ \AA, with 
$t\approx8.6$ \AA. Then, it increases again approaching a finite value 
corresponding to the semiinfinite medium, $t\approx 11.03$ \AA.

\section{Drops and line tension}
\label{ss:dr}
As a next step, we consider  finite systems and, in particular, 
drops of fixed number of atoms $N$. These where already studied by DMC 
techniques in Ref.~\cite{old}. However, computational limitations
allowed only to study small values of $N$. Here, we will take advantage
of the computational feasibility of the Density Functional calculations 
and will extend the analysis to much greater $N$. In this way it is 
possible to study the asymptotic behavior of several quantities which 
characterize the drops. 

In this case, the equation~(\ref{eq:zr-EL}) for the profile 
can be rewriten in the form of a Schr\"odinger-like
equation for $\R$: 
\begin{eqnarray}
   {\cal H}\R \equiv
   -\Dd\left[\nabla^2\R
    - \frac{|\nabla\R|^2}{2\R^2}\right]+
   2b\R + (2+\g)c\R^{1+\g} - 2d\nabla^2\R 
   = \mu\R
  \label{eq:zr-sch}
\end{eqnarray}

As the number of atoms is a well defined $N$, one would need to be very careful
in determining the central density $\rho(0)$ so that the chemical potential
adjusted exactly to $N$. However, it turns out that this equation is more 
efficiently solved by means of the steepest descent method \cite{floca}. 
An initial trial $\rho(r)$ is projected onto the mininum of the functional 
by propagating it in imaginary time. In practice, one chooses a small time 
step $\Delta t$ and iterates the equation 
\begin{equation}
\rho(r,t) \approx \rho(r,t) - \Delta t {\cal H} \rho(r,t)
\label{eq:propa}
\end{equation}
by normalizing $\rho$ to the total number of atoms at each iteration. 
The time step that governs the rate of convergence should be taken 
appropriately small in such a way that Eq.~(\ref{eq:propa}) is a 
valid approximation. Convergence is reached when the
chemical potential has a constant value independent of position.

\begin{figure}[t]
\includegraphics*[width=12cm]{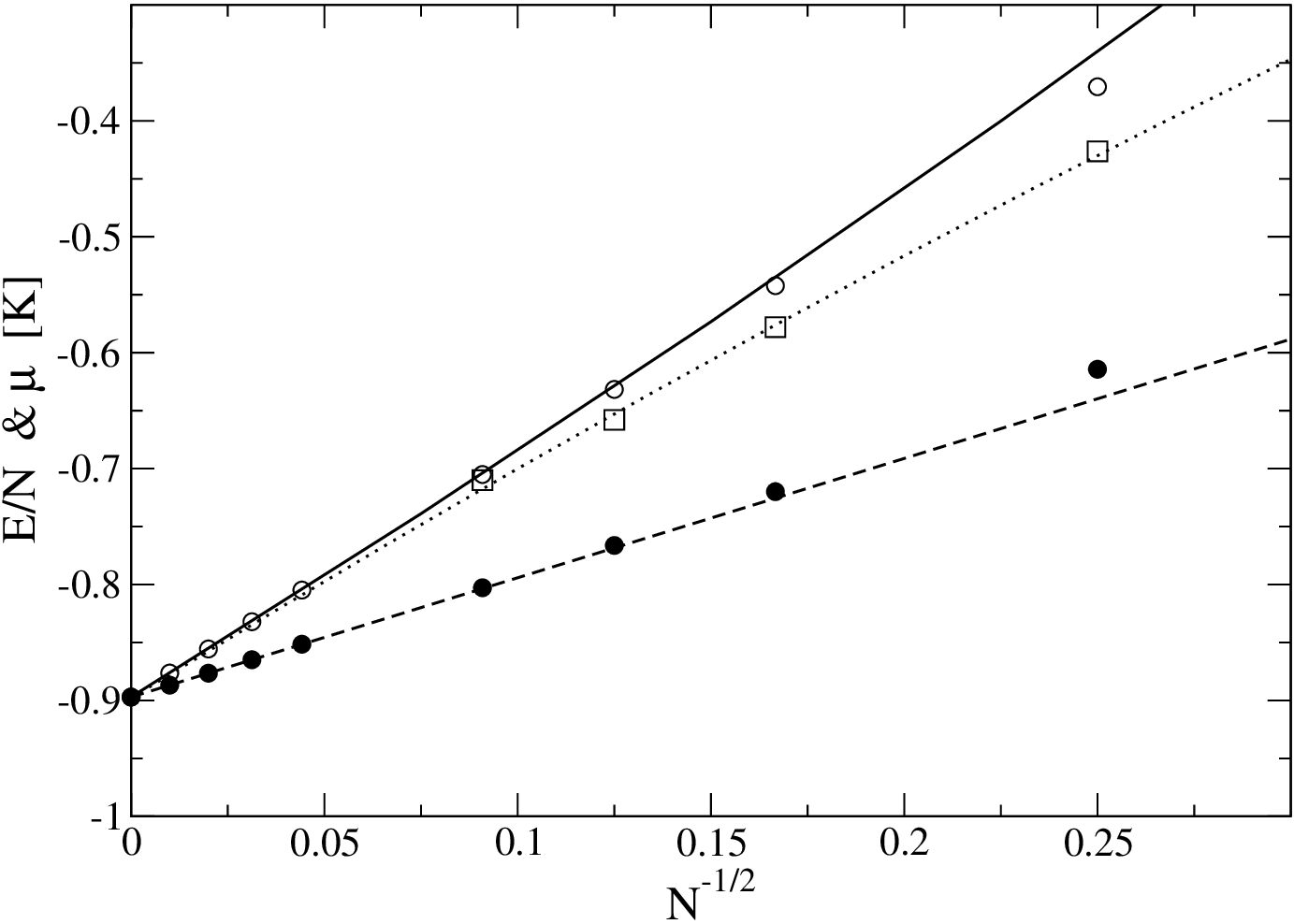}
\caption{Energy per particle  (empty circles) and  chemical potential
(full circles) of $^4$He droplets as a function of $N^{-1/2}$. 
Also shown is a quadratic fit (see text) of the results with 
$N > 516$  (solid line). The straight short-dashed line is obtained when 
the mass formula (with terms up to $z^2$) is used to calculate the chemical 
potential. The empty squares are DMC results from Ref.~\cite{old} and the 
dotted line corresponds to the quadratic fit to these results reported 
in the same reference.} 
\label{fig:enerdrops}
\end{figure}

The energy per particle (empty circles) and the chemical potential 
(full circles) of each drop are reported in Fig. (\ref{fig:enerdrops}) 
as a function of $N^{-1/2}$. Also shown are the DMC results (empty squares) 
and their quadratic fit reported in Ref.
\cite{old}.
The calculated energies of the droplets can be represented very accurately 
with a mass formula of the type
\begin{equation}
  e(N) = \epsilon_b + \epsilon_l z + \epsilon_c z^2 + ...,
\label{eq:massfor}
\end{equation}
with $z = N^{-1/2}$. The two first coefficients of this expansion are the bulk 
energy $\epsilon_b$ and the line energy $\epsilon_l$, out of which the line 
tension $\lambda$ is defined by $2 \pi r_0 \lambda = \epsilon_l$. Here $r_0$ 
is the unit radius, defined as the radius of a disk whose surface is equal 
to the inverse of the equilibrium density of the 2D bulk liquid, i.e, 
$\rho_0 \pi r_0^2 = 1$. Finally, $\epsilon_c$ can be related to the so-called
curvature energy.  Contrary to the DMC calculations where the largest droplet
that we studied had 121 atoms, here we  have considered droplets with up to
 10000 atoms. In this way we can accurately study the behavior of the energy 
per particle for small values of $z$. Doing  a quadratic fit to the 
calculated energies per particle  for $N \ge 512$ and including also the bulk
binding energy for $z=0$ one gets   
\begin{equation}
y = -0.897 + 2.0587 z + 0.66466 z^2 ,
\end{equation}
which is plotted by a solid line in Fig. (\ref{fig:enerdrops}).
One sees that $\epsilon_b$ accurately reproduces the bulk energy per particle, 
which  was used to fix the parameters of the functional.
The value of $\epsilon_l = 2.0587$ corresponds to a line tension 
$\lambda =0.121 K/$\AA, which is the same as  the value of the  line tension 
of the semiinfinite system,  
used to build the density functional. Note that this fit, even if it has been 
calculated for $N \ge 512$, is rather accurate down to $N=36$. 
Obviously, one can not expect a good agreement for $N=16$.
The linear behavior of the chemical potential as a function of $z$ 
is  easy to understand 
using the mass formula (Eq.(\ref{eq:massfor})) and the thermodynamic definition of the chemical potential 
$\mu = \partial E / \partial N$,
where $E$ is the total energy of the droplet. Using this prescription,
 the slope of the chemical potential as a function of $z$ results to be
$\epsilon_l /2$. 
A similar plot for the energy per particle in the three dimensional case
in terms of  $N^{-1/3}$, would provide a behavior
of the chemical potential for large $N$, dominated
also by a linear component with a slope at the origin given by $2 \epsilon_s /3$,
where $\epsilon_s$ is the surface energy 
associated to  three dimensional clusters \cite{str87}.  

Also interesting is the fact that the coefficient of $z^2$ is positive.  
This signe corresponds to the expected  loose of binding energy associated to 
the curvature of the contour of the cluster. This is in contrast with the 
value of $\epsilon_c$ obtained by fiting DMC results as it was done in Ref.
\cite{old} . 
However, in that case the number of particles in the clusters used 
to build the fit was much smaller, being $N=121$ the largest number of particles
and going down to N=8 for the smallest one. In the present fit we have explicitly
avoided the clusters with a small number of particles which can easily distort the 
results, and we have considered only the cases with $N \ge 516$. 

\begin{figure}[t]
\includegraphics*[width=16cm]{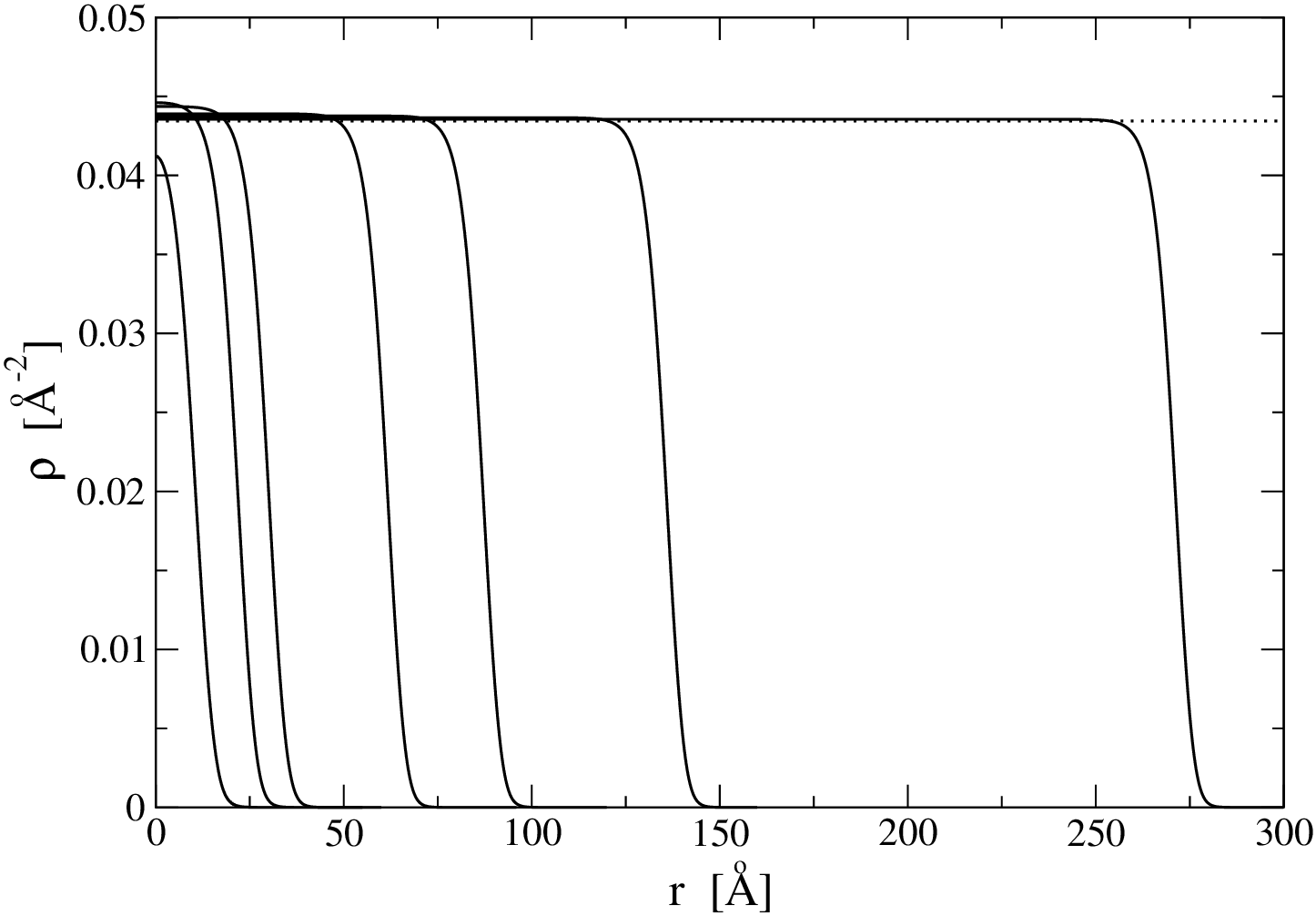}
\caption{Density profiles for $^4$He droplets for $N$=16, 64, 121, 512, 1024, 
2500 and 10000 atoms. 
The continuous lines are generalized Fermi profiles 
(Eq.~\ref{eq:ferprofi}) fitted to the data (see Table \ref{tab:fermi1}). 
The dotted horizontal line indicates the equilibrium density $\rho_0$.} 
\label{fig:prof-drops}
\end{figure}

Next thing to analyze are the density profiles which are reported in 
Fig. ~(\ref{fig:prof-drops}) for different numbers of atoms. Contrary to the
slabs, 
the central density of the droplets can be higher than the saturation density,
which is indicated in the figure by an horizontal line. The profiles are well 
adjusted by a function of the type:
\begin{equation}
\rho(r) = \frac {\rho_f}{\left (1 + e^{\frac {r-R}{c}}\right )^{\nu}}
\label{eq:ferprofi}
\end{equation}
which has an associated central density $\rho(0) = \rho_f /(1+e^{-R/c})^{\nu}$.
The parameters  defining the profiles for the different numbers of atoms 
are provided in Table \ref{tab:fermi1}, together with the thickness and root-mean-square
radius obtained from these fits.

\begin{figure}[t]
\includegraphics*[width=12cm]{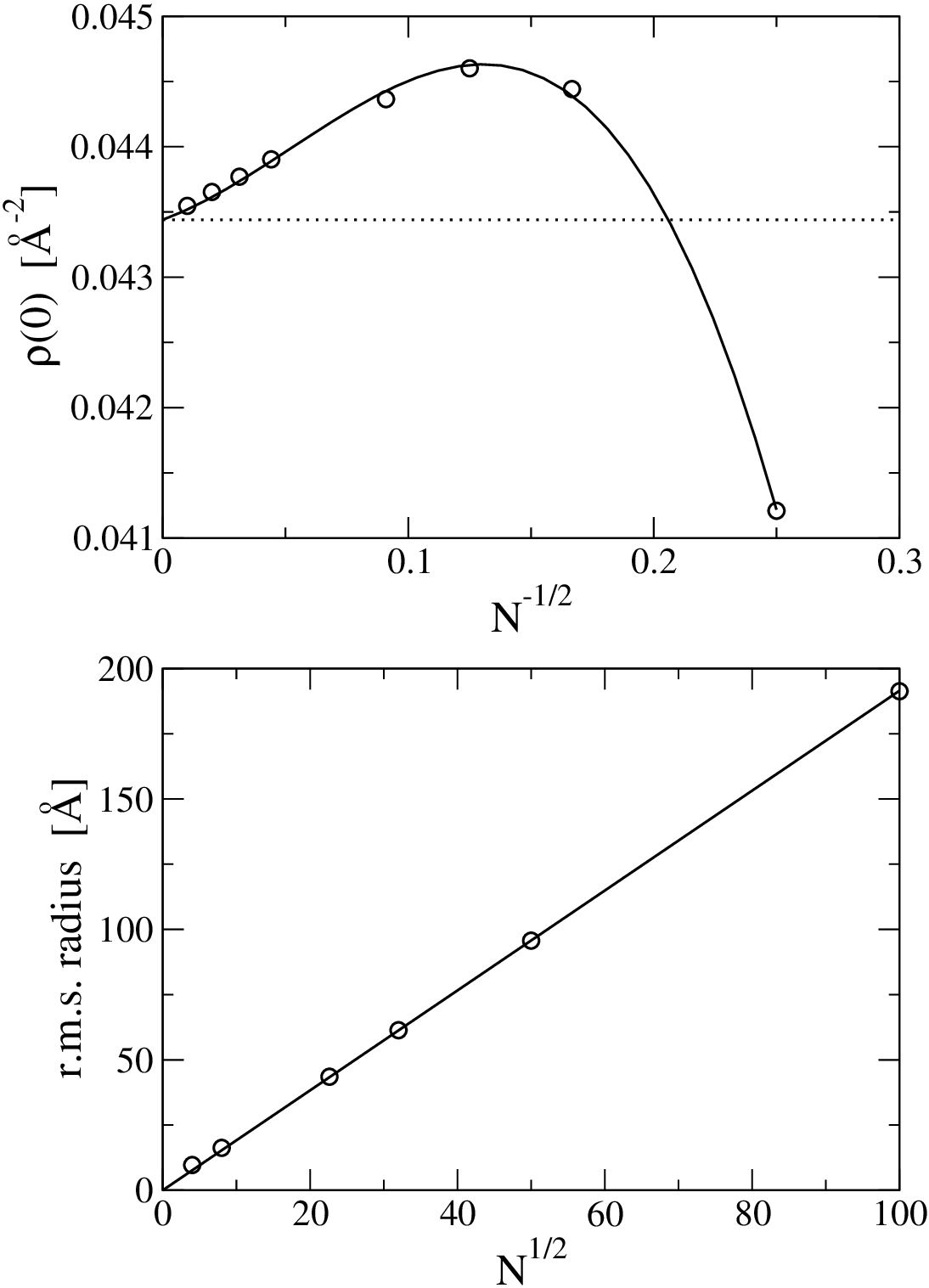}
\caption{(top) Central density of $^4$He droplets as a function of $N^{-1/2}$. 
The empty circles correspond to the results obtained with the zero-range 
density functional, while the line stands for a cubic spline fit to these data.
The dotted horizontal line indicates the saturation density $\rho_0$.
(bottom) Mean square radius of $^4$He droplets as a function of $N^{1/2}$. 
The solid line is a linear fit to the data without independent term.
The empty circles correspond to the results of the zero-range density 
functional.}
\label{fig:central-density}
\end{figure}

The upper part of Fig.~(\ref{fig:central-density}) reports the central 
density of the different droplets as a function of $z$. For large values 
of $N$, the central density
is larger than the saturation density, i.e. the central part of the droplet
is more compressed than the bulk system, which is sometimes referred to as 
a leptodermous behaviour~\cite{str87}.  Of course for $N \rightarrow \infty$ 
the central density tends
to the equilibrium density of the homogeneous liquid. First, the central 
density grows almost linearly with $z$, reaches a maximum, for 
$N \sim 60$, which would correspond to the most compressed droplet and then
decreases. Finally  for $N \leq 25$ the central region of the droplets 
becomes less compressed than the bulk system.

The mean square radius is shown in the lower panel of 
Fig.~(\ref{fig:central-density}) as a function of $N^{1/2}$. 
The expected linear behavior, associated to a constant average density 
\begin{equation}
  \langle  r^2  \rangle ^{1/2} = \frac {1}{\sqrt {2 \pi \rho_0} } N^{1/2}
\end{equation}
is rather apparent. The fit to the calculated values, from $N=16$ to $N=10000$,
provides a $\rho_0 =0.0434$ \AA$^{-2}$ in very good agreement with the 
equilibrium density used to define the parameters of the density functional.

\section{Conclusions}
\label{s:conc}

We have constructed a density functional suitable to study non-homogeneous 
two-dimensional $^4$He systems. First, we have considered two dimensional 
slabs, to study the energy per particle, the central density, radius and density profiles
as a function of the coverage. We have analitically shown  that the extracted  linear tension
from a mass formula adapted to this  type of geometry, is consistent with the value of the linear
tension of the semi-infinite system which was used to build the density functional. 
The thickness and the central density of the slabs approach from below 
the values corresponding to the semi-infinite system while the radius diverges.

We have also studied the energetics and structure of two dimensional clusters. In particular,
we have considered clusters with a very large number of atoms to study  the behavior
of the mass formula and to establish how the system approaches the bulk limit. 
The central density of the clusters when $N \rightarrow \infty$ approaches the 
saturation density from above, and therefore the internal regions of  the clusters 
 are more compressed
than the bulk system while the external regions have densities which would correspond to negative 
pressures or even below the spinodal point for a uniform system. 

The profiles of the clusters are very well fitted by a generalized Fermi function. The 
thickness of the cluster slowly approaches the thicknes of the semi-infinite system, 
as it  also hapens in the case of the slab geometries, but in this case from above.
Finally, we have analyzed the linear behavior of the rms radius of the droplets in terms
of $N^{1/2}$, and recovered the saturation density from the slope of this fit.  
The proposed density functional can be used with a very small computational effort 
for large clusters where the value of $N$ is prohibitive for a Monte Carlo calculation.   

\acknowledgments
This work has been supported by DGICYT (Spain) contracts BFM2002-01868,
FIS2004-00912 and by 
Ge\-ne\-ra\-litat de Catalunya project 2001SGR00064.
J. M. P. acknowledges a
fellowship from the Ge\-ne\-ra\-litat de Catalunya.



\pagebreak

\begin{table}[t]
\begin{center}
\caption{Parameters of the two-dimensional zero-range functional.}
\label{t:func}
\begin{tabular}{cccc}
$b$ [K\AA$^2$] & $c$ [K\AA$^{2(1+\g)}$] & $\g$ & $d$ [K\AA$^4$]\\
\hline
-26.35         & 4.88$\times10^5$       & 3.62 & 359
\end{tabular}
\end{center}
\end{table}

\pagebreak

\begin{table}[t]
\begin{center}
\caption{Parameters of a generalized Fermi-profile fit (Eq.~\ref{eq:ferprofi})
         to the density 
         profiles obtained with the zero-range density functional. 
         All lengths are in \AA\ and $\R_f$ is in \AA$^{-2}$. 
	 The parameter $\nu$ is adimensional.}
\label{tab:fermi1}
\vspace{0.3cm}
\begin{tabular}{c|cccc|cc}
\hline
$N$ & $\R_f$ & $R$ & $c$   & $\nu$ & $t$   & $\langle r^2\rangle^{1/2}$ \\ 
\hline
16 & 0.04321 & 13.2718 & 3.22067 & 2.13417 & 11.746 &  9.66\\
36 & 0.04494 & 19.1852 & 3.22054 & 2.37516 & 11.548 &  12.793    \\
64 & 0.04974 & 24.6826 & 3.16845 & 2.37072 & 11.364 & 16.28\\
121& 0.04441 & 32.8314 & 3.12302 & 2.33636 & 11.226 & 21.72\\
512& 0.04392 & 64.4245 & 3.08867 & 2.32368 & 11.112 & 43.533\\
1024& 0.04378& 89.8497 & 3.08584 & 2.32997 & 11.097 & 61.342 \\
2500& 0.04366& 138.628 & 3.08552 & 2.33826 & 11.090 & 95.678 \\
10000& 0.04355& 274.026 & 3.08653 & 2.3467 & 11.088 & 191.276 \\
\hline
\end{tabular}
\end{center}
\end{table}


\pagebreak

\pagebreak

\pagebreak
\pagebreak

\pagebreak

\end{document}